\documentclass[runningheads]{llncs}
\usepackage[T1]{fontenc}
\usepackage{graphicx}
\usepackage{wrapfig}
\usepackage{multirow}%
\usepackage{amsmath,amsfonts}
\usepackage{xcolor}
\usepackage{comment}
\usepackage{booktabs}
\usepackage[export]{adjustbox}
\usepackage{placeins}
\usepackage{pdfcomment}

\usepackage[labelformat=simple]{subfig}

\usepackage[labelfont=bf]{caption} 

\begin{document}
\title{Confident Teacher, Confident Student? A Novel User Study Design for Investigating the Didactic Potential of Explanations and their Impact on Uncertainty}
\titlerunning{Confident Teacher, Confident Student?}
%
\author{Teodor Chiaburu\inst{1}\orcidID{0009-0009-5336-2455} \and
Frank Haußer\inst{1}\orcidID{0000-0002-8060-8897} \and
Felix Bießmann\inst{1,2}\orcidID{0000-0002-3422-1026}}
\authorrunning{T. Chiaburu et al.}
%
\institute{Berliner Hochschule für Technik, Luxemburger Str. 10, 13353 Berlin, Germany
\email{\{chiaburu.teodor,frank.hausser,felix.biessmann\}@bht-berlin.de}
\and
Einstein Center Digital Future, Wilhelmstraße 67, 10117 Berlin, Germany}
\maketitle              
\begin{abstract}
Evaluating the quality of explanations in Explainable Artificial Intelligence (XAI) is to this day a challenging problem, with ongoing debate in the research community. While some advocate for establishing standardized offline metrics, others emphasize the importance of human-in-the-loop (HIL) evaluation. Here we propose an  experimental design to evaluate the potential of XAI in human-AI collaborative settings as well as the potential of XAI for didactics. In a user study with 1200 participants we investigate the impact of explanations on human performance on a challenging visual  task - annotation of biological species in complex taxonomies. Our results demonstrate the potential of XAI in complex visual annotation tasks: users become more accurate in their annotations and demonstrate less uncertainty with AI assistance. The increase in accuracy was, however, not significantly different when users were shown the mere prediction of the model compared to when also providing an explanation. We also find negative effects of explanations: users tend to replicate the model's predictions more often when shown explanations, even when those predictions are wrong. When evaluating the didactic effects of explanations in collaborative human-AI settings, we find that users' annotations are not significantly better after performing annotation with AI assistance. This suggests that explanations in visual human-AI collaboration do not appear to induce lasting learning effects. 
All code and experimental data can be found in our GitHub repository: \url{https://github.com/TeodorChiaburu/beexplainable}.

\end{abstract}

\section{Introduction}\label{sec:intro}

XAI strives to bridge the gap between complex AI models and human users by providing explanations for the models' decisions. However, evaluating the effectiveness of XAI methods remains a challenging question\cite{anecdotal,hoffman2019metrics,ma2024openhexai,info14080469,Schuff2022ChallengesIE,mohseni2020multidisciplinary}. Currently, two primary methodological approaches dominate the XAI evaluation landscape: \textbf{fully automated evaluation metrics} and \textbf{HIL approaches}. While the former strategy offers an objective and scalable solution \cite{hedstrom2023quantus,autoxai,herman2019promise}, it often lacks the richness of human understanding.

HIL experiments are supposed to provide a powerful alternative by directly assessing user interaction with explanations. While human comprehension is the ultimate goal of explainability, evaluating model explanations through human subject studies presents a significant challenge. These studies necessitate rigorous experimental design and can be resource-intensive in terms of time, money or logistics \cite{leavitt2020falsifiable}. Nonetheless, due to the inherently human-centric nature of explainability, a substantial body of literature advocates for HIL experiments as a means to assess explanation quality \cite{doshi_velez_towards_2017,Kim2022HIVE,colinWhatCannotPredict,kindermans_reliability_2017,adebayo_sanity_2020,mei2023users}. 

Several aspects of understanding the effect of explanations are of interest and worth exploring for a  given XAI method, as discussed in the following section. While allowing for analyzing some of these points, the HIL framework we propose in this work enables practitioners to also measure the didactic potential of explanations, as well as correlations between the subjects' and the machine's uncertainty. To the best of our knowledge, these aspects have yet to be investigated in the current XAI literature.

\section{Related Work}\label{sec:rel_work}

The definition of a "good" explanation remains an ongoing debate, leading to a diverse range of HIL experiments focusing on various aspects and metrics. Comprehensive reviews of this field can be found in \cite{Ruong2024,exp_matters,Schwalbe_2023}.

Perhaps the most extensively explored facet of XAI research concerns the influence of explanations on \textbf{human performance}. Do explanations actually help users perform better? This can be investigated through two main lenses: \textit{human-AI collaboration} and \textit{knowledge transfer}. 

In human-AI collaboration tasks, explanations can enhance user performance by effectively communicating the AI's reasoning. This allows users to better utilize the AI's suggestions, adjust their own decisions alongside the AI's input and potentially improve overall task efficiency through smoother collaboration. HIL experiments can shed light on these aspects by observing user behavior in such team-settings. The literature abounds in studies investigating scenarios where users work with AI assistants to solve various tasks. These studies often demonstrate that AI assistance improves human performance in tasks like sentiment analysis \cite{schmidt2019quantifying}, poisonous mushroom classification \cite{xai_mushrooms}, insulin dosage decisions for virtual patients \cite{exp_insulin} or prostate cancer classification in MRI scans \cite{radiology}.

In terms of long-term benefits of explanations, the question is whether explanations can empower users to learn from the AI and improve their independent performance on future tasks. Effective explanations might enhance user understanding of relevant rules and patterns used by the AI. This understanding could then translate to improved task execution without AI assistance, potentially promoting long-term learning that can be utilized for various tasks beyond the specific context of the AI system. HIL experiments can be designed to assess whether explanations facilitate such knowledge transfer, basically crystallizing a pedagogic effect of XAI. As previously stated, to the best of our knowledge, there are no studies available at the time of writing this paper, that deal with the didactics of explanations. Our proposed HIL framework attempts to cover this gap.

Still related to performance, the aspect of \textbf{simulatability} is also frequently investigated in HIL studies. Simulatability refers to whether users can understand and replicate the model's reasoning based on the explanations provided. This topic is extensively discussed in \cite{hase2020evaluating} and \cite{doshi_velez_towards_2017}, where the authors distinguish between "forward simulation" (users predict the model's output for a given input and explanation) and "counterfactual simulation" (given an input, a model's output and an explanation, users predict the model's output for a perturbation of that input).

Another key factor is the effect of explanations on \textbf{trust}, namely how they calibrate user trust in the model's predictions or how they can mitigate the issue of blind trust in situations where the model might be less confident. For instance, in \cite{Kim2022HIVE} participants are shown the prediction of a computer vision model along with an explanation in the format "Class A because ..." and are asked how confident they are in the model's prediction. On a different note, \cite{biessmann_turing_2021} investigate in a Turing test inspired approach whether subjects are able to distinguish between human-generated and AI-generated explanations and argue that such a quantitative metric, employed alongside trust calibration techniques, would offer valuable insights into how intuitive an explanation is. Clear, comprehensive and accurate explanations can help users assess the AI system's competence and expertise. If explanations effectively reflect the model's reasoning process, users are more likely to believe the AI is knowledgeable and capable of assisting with the task at hand. This fosters trust in the AI's ability to provide accurate suggestions and recommendations. Explanations can also mitigate the risk of blind trust. By highlighting the AI's limitations and uncertainties, explanations can encourage users to approach the AI's suggestions with a healthy dose of skepticism.  This allows them to maintain an appropriate level of critical thinking and intervene when necessary without entirely disregarding the AI's input. Ultimately, explanations should strive to create a balance between trust and critical engagement with the AI system.

A natural question stems from the issue of trustworthiness: does \textbf{user uncertainty} align with \textbf{model uncertainty} in XAI contexts? This is particularly important for building trust in situations where the model might be less confident in its predictions \cite{chiaburu_uncertainty}. Again, as far as we are aware, this aspect has yet to be investigated in HIL studies in the available literature so far. 

The perceived \textbf{cognitive effort} required to understand explanations is another important consideration. Several studies highlight that explanations may not always have a solely positive impact and could even yield negative effects on human subjects during cognitive tasks \cite{schmidt2020transparency,alufaisan2020does,david2021explainable,poursabzi-sangdehManipulatingMeasuringModel2018}. These studies often attribute performance declines in cognitive tasks to the increased cognitive load imposed by explanations \cite{poursabzi-sangdehManipulatingMeasuringModel2018}. 

Lastly, XAI research also considers the perceived \textbf{usefulness} and perceived \textbf{ease of use} of explanations, e.g., \cite{exp_hate_speech}. This focus acknowledges that explanations must not only be understandable, but also practically valuable for users. If explanations are deemed unhelpful or difficult to grasp, they are unlikely to enhance user experience or performance.

The following sections present our proposed HIL approach and the methods of investigating the didactic effect of explanations, the human-machine collaboration, the degree of trust users have in AI, as well as the relation between the users' and the model's uncertainty within this framework.

\section{Dataset and Classification Problem}\label{sec:data}

For the experiments described in this work, a subset of the iNaturalist dataset \cite{iNat} was used. A dataset of wild bee images was constructed by scraping 30k images from the online database \url{https://www.inaturalist.org/}. These images depicted the top 25 most frequent wild bee species native to Germany within their natural habitats. Following preliminary experiments, the dataset was refined to focus on three particularly challenging and frequently confused species: Andrena bicolor/flavipes/fulva (see Fig. \ref{fig:protos}). This refinement resulted in a final subset containing 657 wild bee images. For more details on the scraping process, data split and annotation, as well as training of the model - a ResNet50 \cite{resnet} - please consult \cite{chiaburu2024copronn} and our repository - \url{https://github.com/TeodorChiaburu/beexplainable}.

\def\wPlotsSpecif{0.32}
\begin{figure*}[h!]
    \centering
    \subfloat[Andrena bicolor]{\includegraphics[width=\wPlotsSpecif\textwidth]{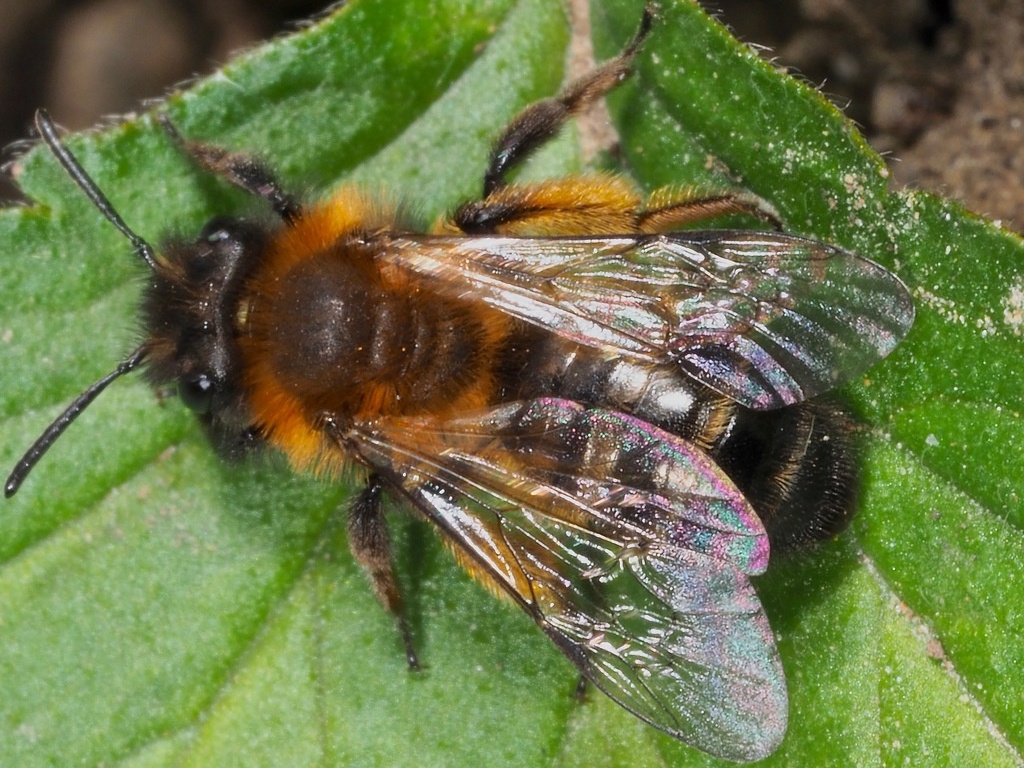}}\hfill
    \subfloat[Andrena flavipes]{\includegraphics[width=\wPlotsSpecif\textwidth]{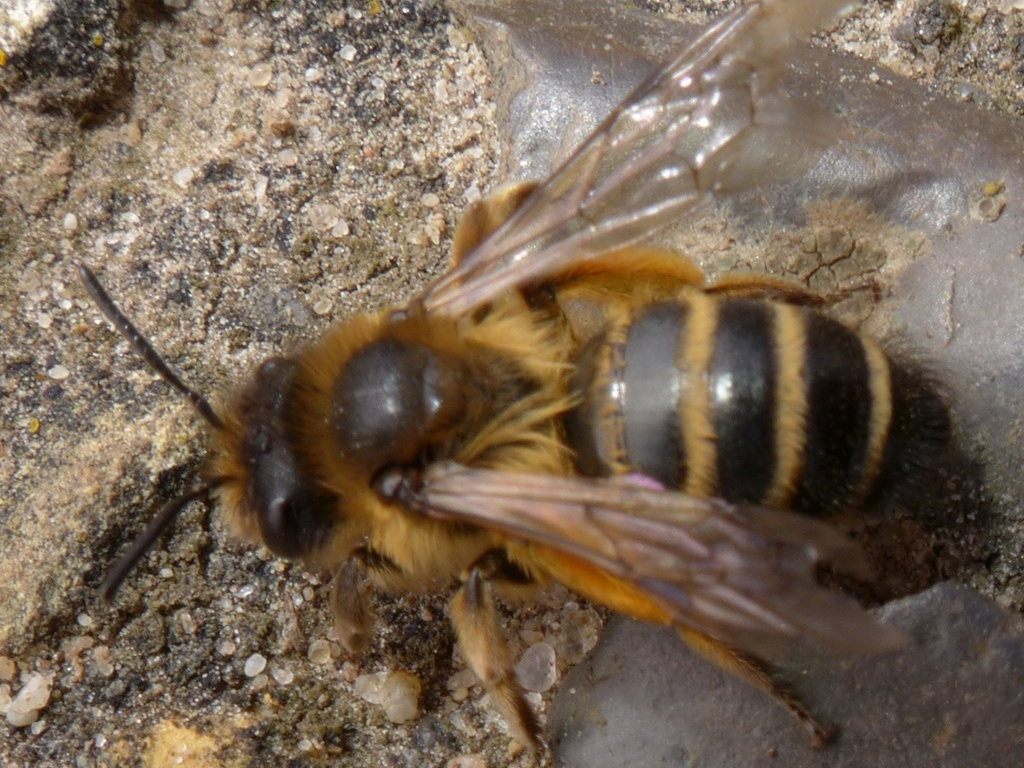}}\hfill
    \subfloat[Andrena fulva]{\includegraphics[width=\wPlotsSpecif\textwidth]{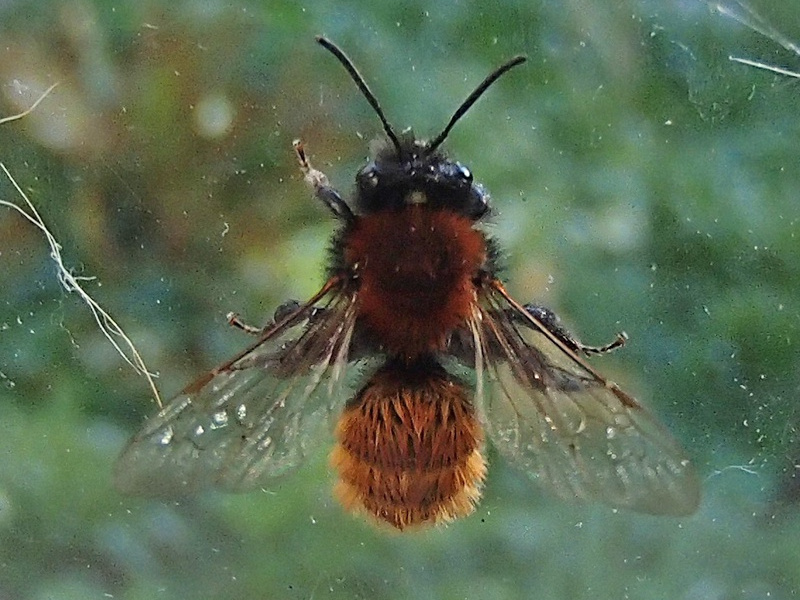}}\hfill
    \hfill
    \caption{Prototypical examples of the three wild bee species used for our HIL experiment. These examples were also shown to the participants in the instructions at the beginning of the trial. The difficulty in distinguishing the three species from one another consists in morphological features present on the bees' thorax and abdomen: A. bicolor has a fuzzy orange thorax and a shiny brown abdomen; A. flavipes has a fuzzy brown thorax and shiny brown abdomen; A. fulva's thorax and abdomen are both fuzzy orange.}
    \label{fig:protos}
\end{figure*}

\section{Experimental Design}\label{sec:exp_design}


\begin{figure}[htbp]
    \hspace*{-3.5cm}
    \includegraphics[scale=0.75]{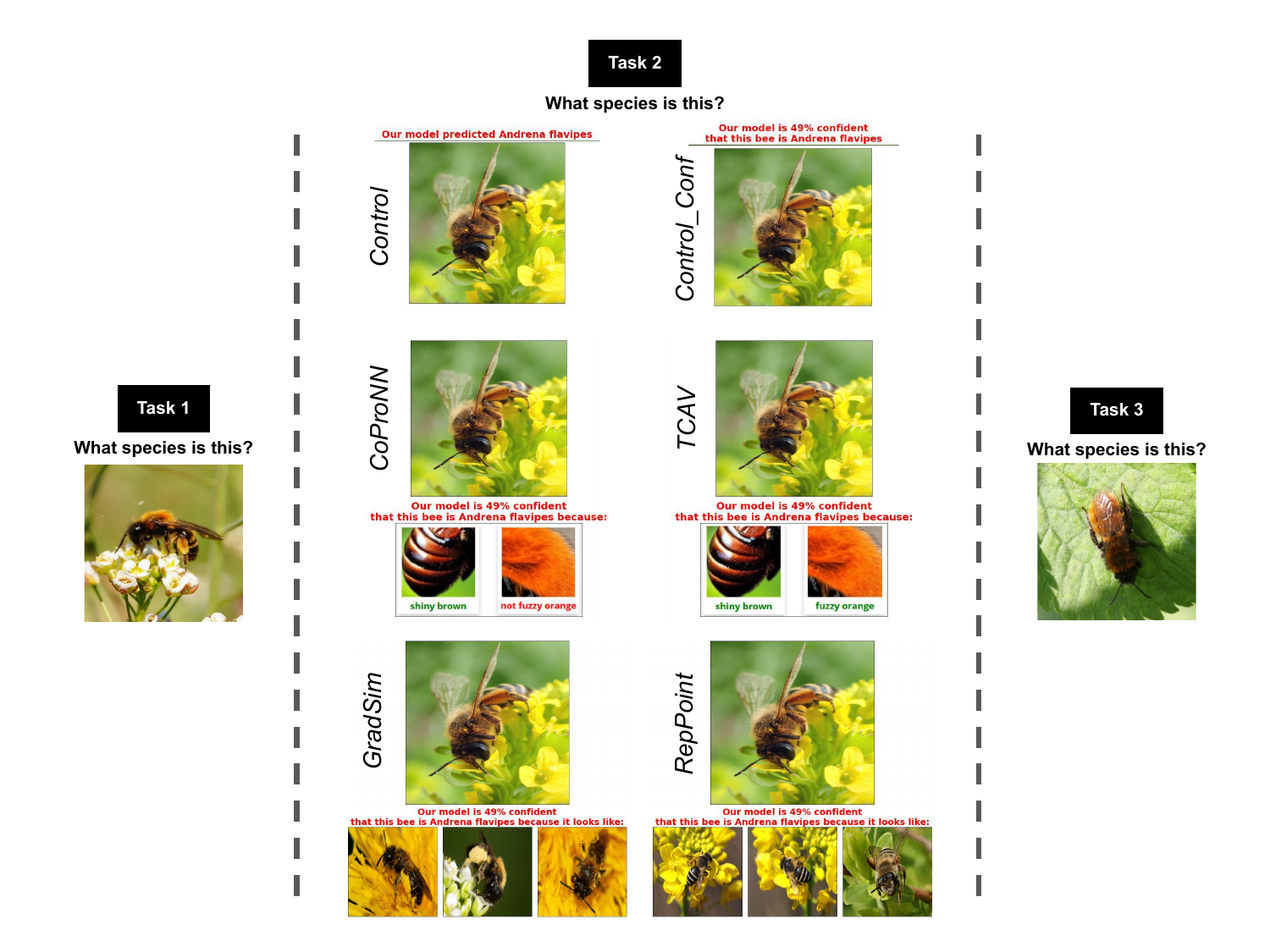}
    \caption{Experimental Design. From left to right: Task 1 - users classify images on their own; Task 2 - further AI assistance is provided, presented differently depending on the assigned group; Task 3 - users again classify images alone. 
    }
    \label{fig:exp_design}
\end{figure}

Our experimental setup consists of three tasks - see Figure \ref{fig:exp_design}. The users are shown images of wild bees and are required to recognize the three species depicted in Fig. \ref{fig:protos}. In Task 1, subjects are left to assign on their own the correct label to the images they see. In Task 2, they are aided by a computer vision model trained to recognize wild bees. In the third and final task, the photos are again shown without any AI hints. Each task comprises 10 images. The reason for adding a second 'control task' (Task 3) after the AI-assisted Task 2 is to enable the investigation of the potential didactic effect that explanations may have. We hypothesize that, if explanations are able in a pedagogical sense to teach laypersons relevant classification rules, then the users' accuracy in Task 3 should be higher than in Task 1, when participants were just getting acquainted with the problem. 

The 30 images that every participant sees throughout the trial are randomly drawn from a pool of 45 samples (selected from the test set). The pool is a mixture of 'easy' and 'hard' examples with respect to the model's confidence in classifying those samples. We label an image as 'hard' if the model assigned a true class probability below $80\%$. Inherently, some of the hard samples were misclassified by the network. When compiling the set of 30 images shown to a subject, we ensured that each task contains 5 easy samples and 5 hard ones.

The participants were informed that the data gathered would solely be used for research purposes. Their identities were anonymous and before starting they were given a detailed description of what they were required to do. For each possible class, a representative example was shown in the introduction, which they could refer back to any time during the trial. The experiment was approved by our institutional research board. 

Users were divided into 6 groups that differed from one another in the type of AI hint revealed in the second task (see Figure \ref{fig:exp_design}):

\begin{enumerate}
    \item \textit{Control Group}: the AI hint consists solely of the model's predicted class (which can be wrong)
    \item \textit{Control-Confidence Group}: the model's prediction is accompanied by the corresponding Softmax probability (model confidence)
    \item \textit{Concepts-CoProNN Group}: the model's prediction and confidence are shown together with an explanation computed by the concept-based XAI method CoProNN \cite{chiaburu2024copronn}. The explanation is visualized in a 'traffic-lights' format, where a representative patch of the concepts learned by the XAI method is marked as relevant (green) or not (red).
    \item \textit{Concepts-TCAV Group}: the model's prediction and confidence are shown together with an explanation computed by the concept-based XAI method TCAV \cite{tcav}. The explanation modality is the same as for CoProNN.
    \item \textit{Examples-GradSim Group}: the model's prediction and confidence are shown together with an explanation computed by the example-based method Gradient Similarity \cite{charpiat2021input}. The explanation is visualized in the form of the top 3 most similar samples from the training set that were classified similarly.
    \item \textit{Examples-RepPoint Group}: the model's prediction and confidence are shown together with an explanation computed by the example-based method Representer Point Selection \cite{yeh2018representer}. The explanation is visualized the same way as for Gradient Similarity.
\end{enumerate}

We carried out our experiment on the crowdsourcing platform Toloka (\url{https://toloka.ai/}). For each of the above mentioned groups, 200 users were accepted (not counting automatically discarded suspicious bot accounts that complete the whole experiment in a matter of seconds). At the end of the experiment, a separate outlier detection would be applied to every group, filtering out submissions with less than 3 correct answers in any of the three tasks. The number 3 corresponds to the 20\% quantile and was determined based on a pilot study. The subjects were selected from the top 50\% of English-speaking Tolokers and remunerated for their participation with 0.06\$ per task suite.

Before conducting this experiment at larger scale, we designed and performed a smaller pilot study with five wild bee species and only 80 participants divided into two groups: \textit{Control} and \textit{CoProNN}. The experiment was deployed as a jsPsych app \cite{jspsych}. A demo is still freely available online for the CoProNN group at \url{https://hgyl4wmb2l.cognition.run}. For more details about the pilot study, please refer to \cite{chiaburu2024copronn} and our repository.

The following section discusses the results of our experiment. As a forenote regarding the quantification of the user-related and model-related metrics: we report 
the entropy\footnote{\url{https://docs.scipy.org/doc/scipy/reference/generated/scipy.stats.entropy.html}} of the Softmax-normalized probability vector output by our model for every image as the \textit{model's uncertainty} and the entropy over the Softmax-normalized vector of the users' submitted answers for each image as the \textit{users' uncertainty}. We acknowledge that more sophisticated metrics exist for quantifying epistemic uncertainty within model predictions, see e.g. \cite{epistemic_uncert_principles}. However, for the purposes of this study, a proxy measure as defined above was deemed sufficient.

\section{Results and Discussion}\label{sec:res}

\begin{figure}[h]
    \centering
    \includegraphics[width=0.85\linewidth]{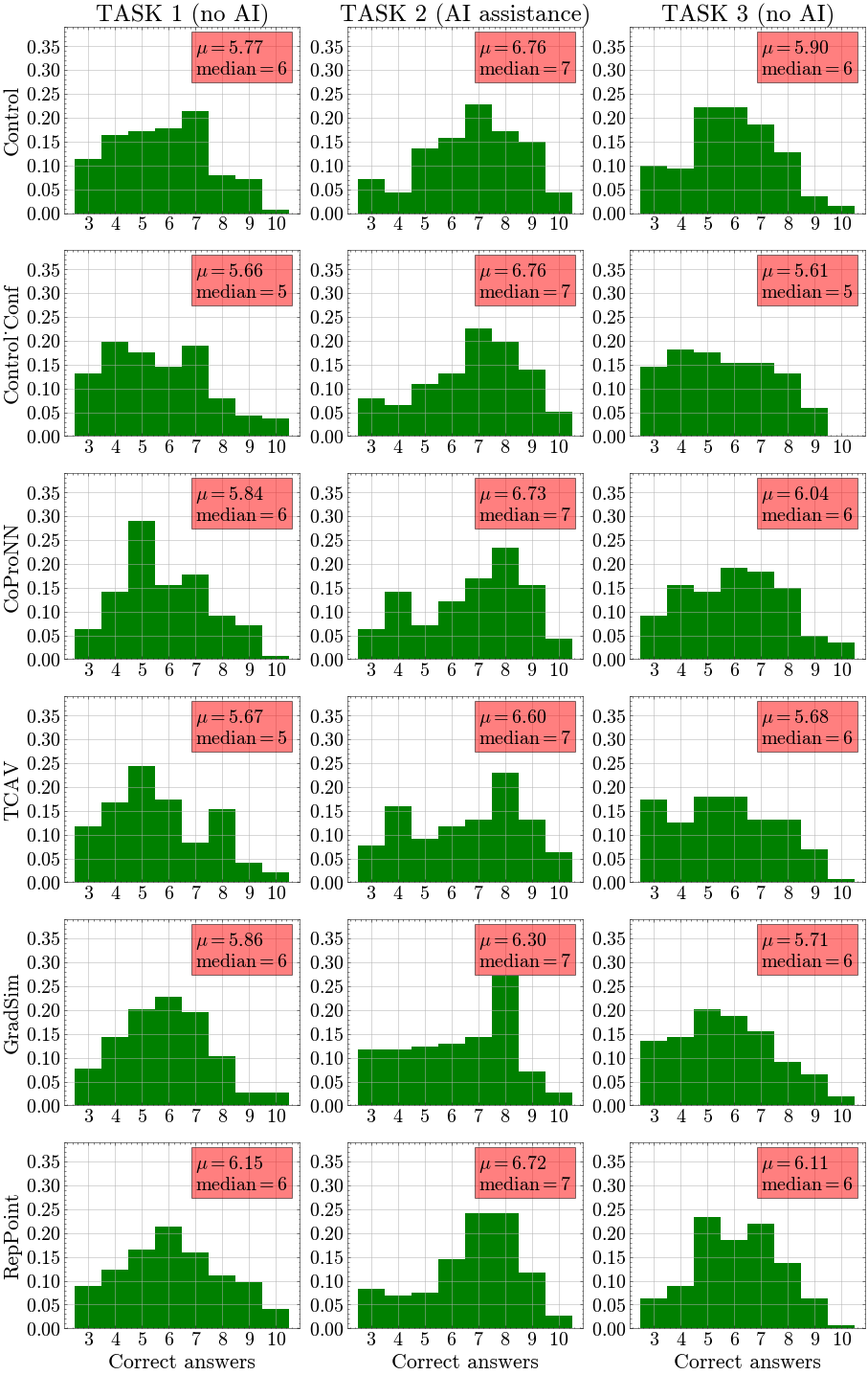}
    \caption{AI assistance improves users' performance, but showing only AI predictions helps as much as showing explanations. Throughout all the 6 human-AI collaboration conditions we notice higher user accuracies in Task 2 (where AI hints were provided) in the mean and median, as opposed to Tasks 1 and 3 (both without AI assistance). In Task 3 participants are not performing any better than in Task 1, suggesting that there is no substantial didactic effect derived from the assistive AI in Task 2. Moreover, there is also no noticeable difference between the two control groups and the four explanation groups. This suggests that explanations were not more effective than simply reporting the model's prediction.
    }
    \label{fig:user_acc_hist}
\end{figure}

We summarize below the insights we gained from our user study. Figure \ref{fig:user_acc_hist} offers a broad overview of the subjects' performance throughout the three tasks and the six groups, while the correlation plots in Figures \ref{fig:acc_uncert} and \ref{fig:uncert_uncert} display the relation between the model's and the users' performance.

\FloatBarrier

\subsection{No Observable Didactic Effects Detected}

Our experiment did not reveal any considerable evidence of long-term learning or knowledge transfer from explanations to independent task performance. This is indicated by the comparable user accuracy observed in Task 3 (where users classified images on their own again) compared to Task 1 (initial independent classification). Across all six groups, both average and median accuracy scores in Task 3 remained similar to those in Task 1 (Figure \ref{fig:user_acc_hist}). This suggests that while explanations may improve performance during collaboration with the AI (as shown in Subsection \ref{subsec:hum_perf}), they may not necessarily equip users with the ability to retain that knowledge and apply it to solve similar tasks independently over time. Tiredness and cognitive load may also play a role once users arrive at the final Task 3.

\subsection{Users' Uncertainty Decreases when Collaborating with an AI Assistant}\label{subsec:hum_uncert}

Our study also revealed a positive impact on user confidence when collaborating with an AI assistant. This is reflected in the user uncertainty levels observed across the different tasks and computed as described at the end of Section \ref{sec:exp_design}. User uncertainty in Task 2, where participants received help from our model, was considerably lower compared to the uncertainty levels observed in Tasks 1 and 3 (Table \ref{tab:user_tab} and Figure \ref{fig:uncert_uncert}). This suggests that hints provided by the AI assistant helped users feel more certain about their classifications in Task 2 and allowed them to approach the task with greater confidence.

\subsection{Human Performance Improves when Collaborating with an AI Assistant}\label{subsec:hum_perf}

Our findings demonstrate that human performance improves when collaborating with an AI assistant. Across all user groups participating in Task 2, both average and median user accuracy scores are consistently higher compared to Tasks 1 and 3 (as illustrated in Figure \ref{fig:user_acc_hist}). Overall, user performance in Task 2 is generally superior to that observed in Tasks 1 and 3, particularly for samples where the model exhibited a high degree of certainty (Figure \ref{fig:acc_uncert}). This indicates that explanations and AI assistance were most beneficial for tasks where the model 
was most certain, potentially aiding users in making more accurate classifications.

\subsection{Limited Impact of Explanation Type on Task Performance}

While Subsections \ref{subsec:hum_perf} and \ref{subsec:hum_uncert} highlighted the overall benefits of collaboration with an AI assistant in Task 2, user performance within this task did not exhibit noticeable differences across the six groups (as depicted in Figure \ref{fig:user_acc_hist}). This suggests that, in the context of our experiment, the specific format or level of detail provided in the explanations (concept-based or example-based) did not have a substantial impact on user accuracy when compared to the two Control groups that received only the model's prediction (with and without model confidence).

\def\wPlotsSpecif{0.75}
\begin{figure*}[h!]
    \centering
    \subfloat[Task 2 (AI Assistance)]{\includegraphics[width=\wPlotsSpecif\textwidth]{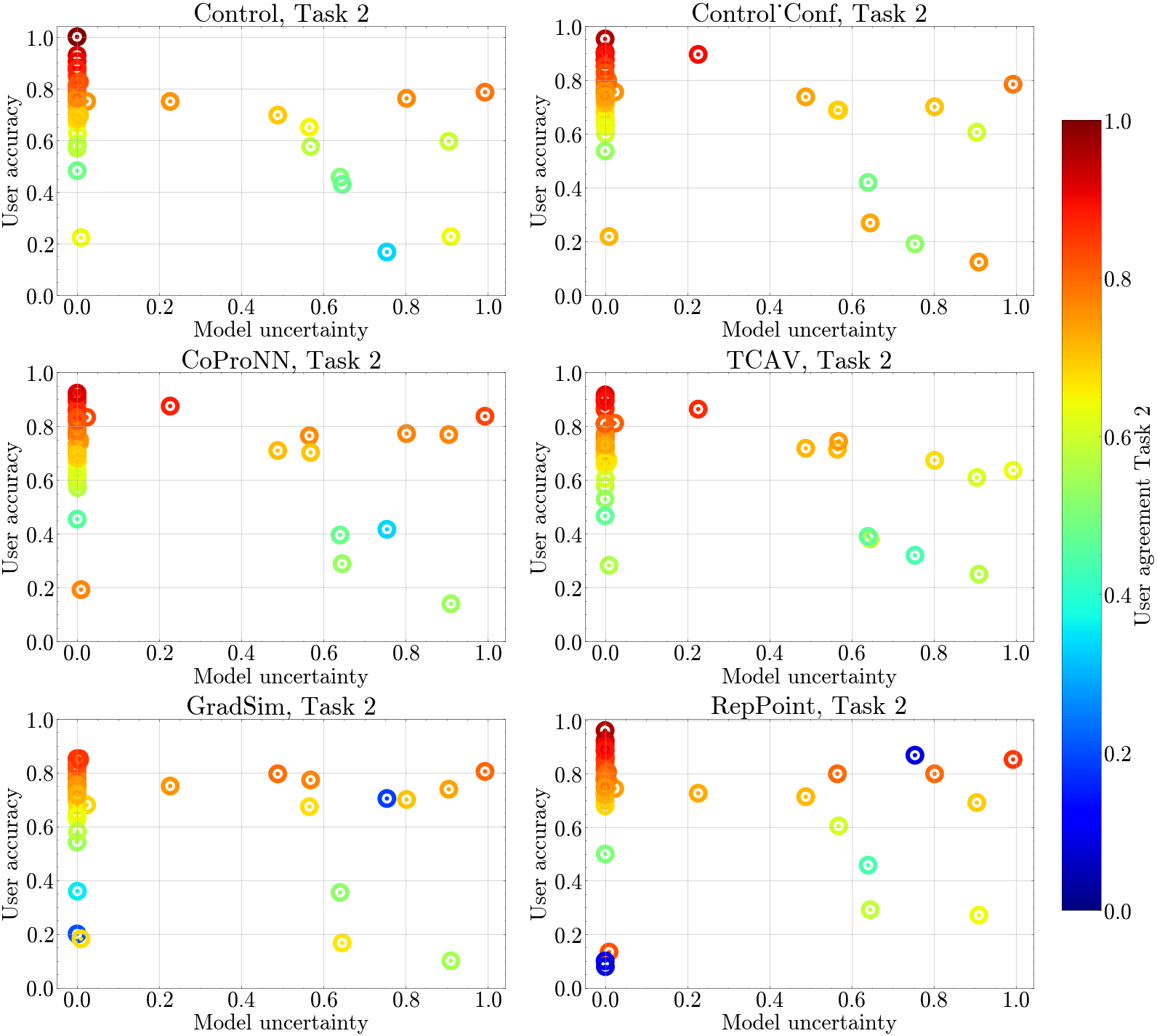}}\hfill
    \centering
    \subfloat[Tasks 1 and 3 (no AI Assistance)]{\includegraphics[width=\wPlotsSpecif\textwidth]{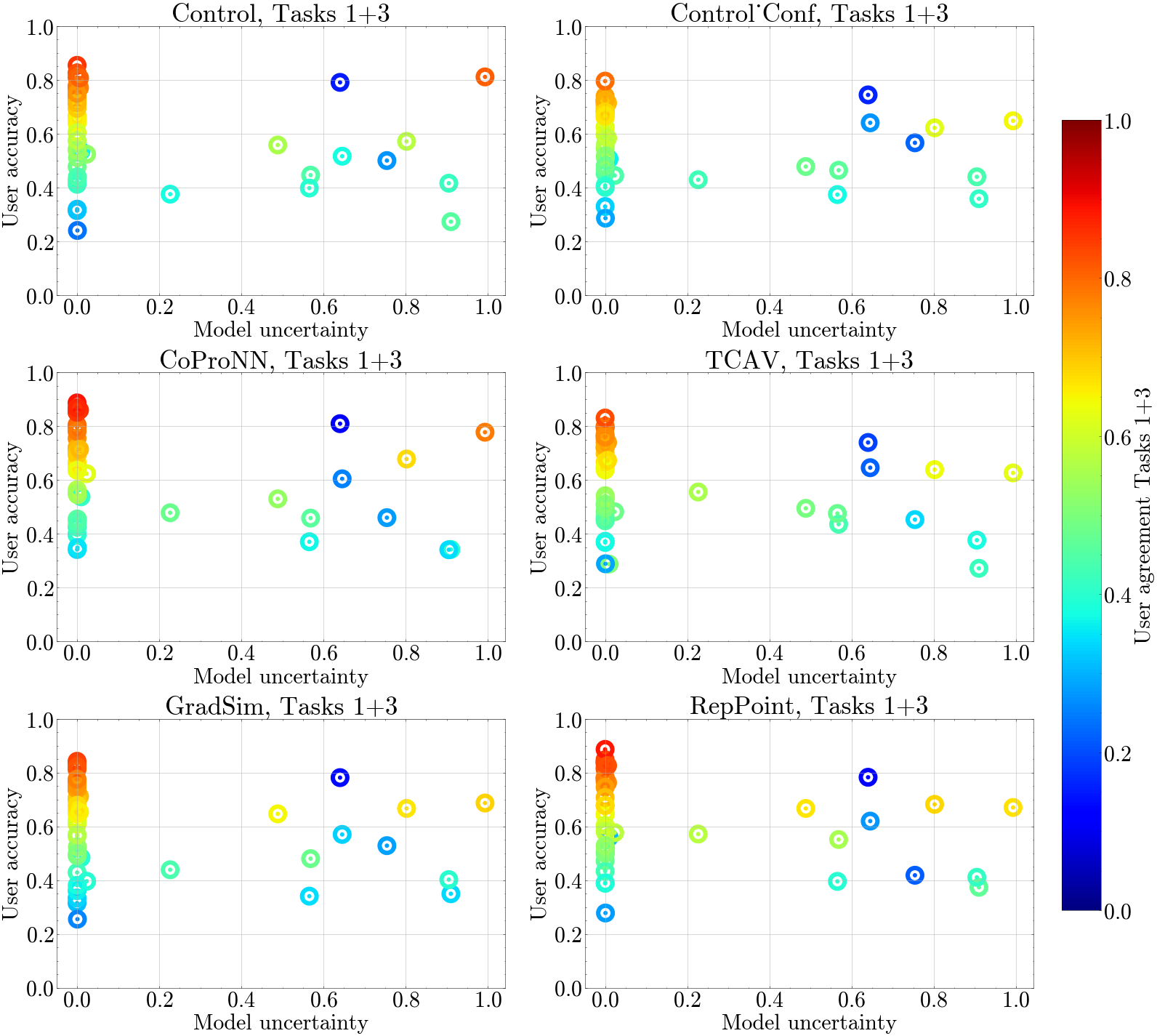}}\hfill
    \caption{Explanations improve user accuracy in human-AI collaboration (Task 2). Every dot represents one of the 45 images with corresponding user metrics averaged over all users' responses. \textbf{(a) Task 2 (with AI assistance)}: Throughout all groups, samples classified with high certainty by the model were also classified with high accuracy by users. These samples are also associated with high acceptance rate for the AI suggestion. \textbf{(b) Tasks 1+3 (without AI assistance)}: User accuracy is lower than in Task 2, regardless of whether the model was certain or not. The user-AI agreement rate is also notably lower.}
    \label{fig:acc_uncert}
\end{figure*}

\def\wPlotsSpecif{0.75}
\begin{figure*}[h!]
    \centering
    \subfloat[Task 2 (AI Assistance)]{\includegraphics[width=\wPlotsSpecif\textwidth]{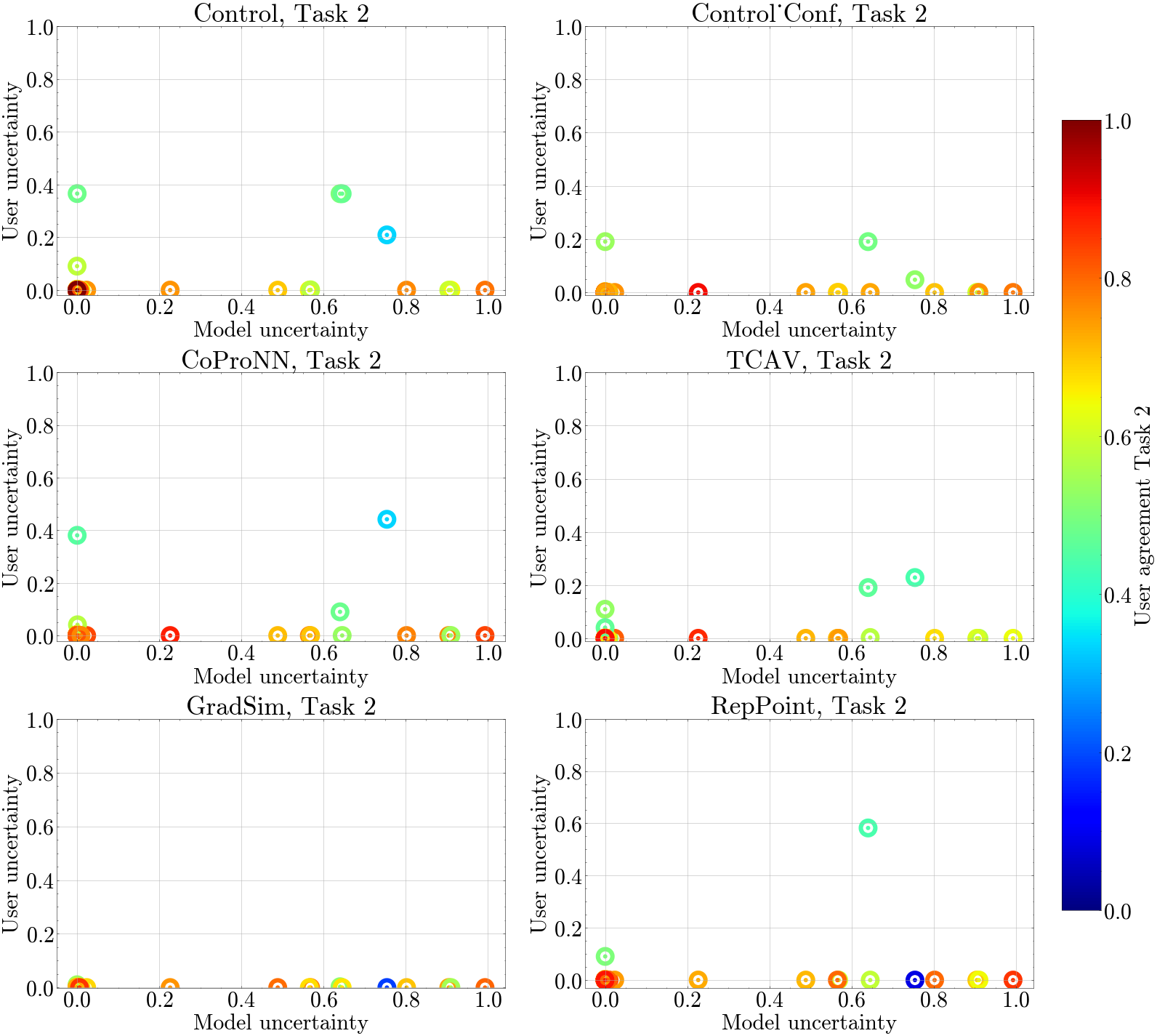}}\hfill
    \centering
    \subfloat[Tasks 1 and 3 (no AI Assistance)]{\includegraphics[width=\wPlotsSpecif\textwidth]{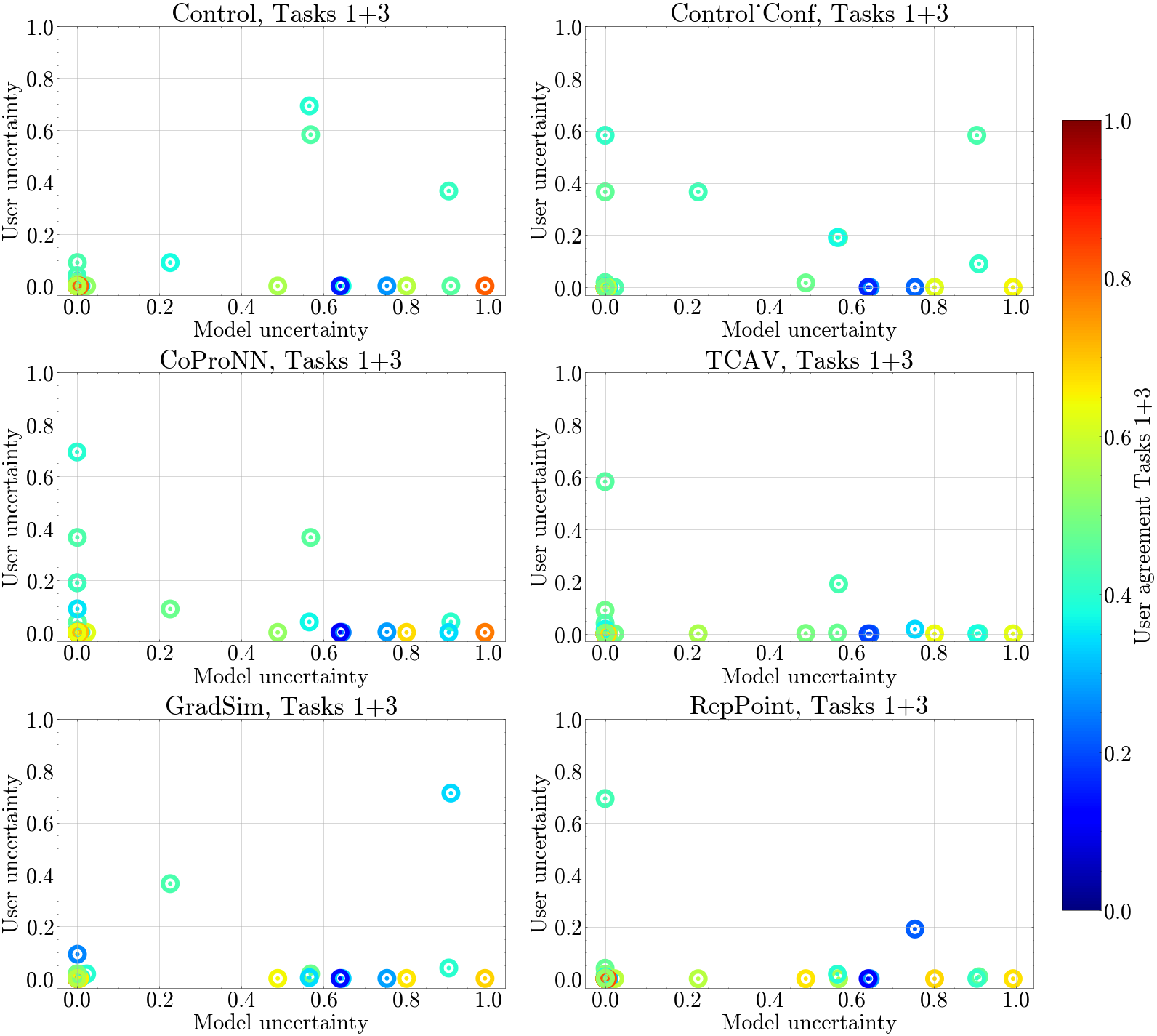}}\hfill
    \caption{Uncertainty of users' responses decreases when explanations are shown (Task 2) compared to no AI assistance (Task 1+3). Every dot represents one of the 45 images with corresponding user metrics averaged over all users' responses. \textbf{(a) Task 2 (with AI assistance)}: Independent of model uncertainty, the users' uncertainty is near 0 for most samples. User-AI agreement is also high with these samples. \textbf{(b) Tasks 1+3 (without AI assistance)}: Users are less certain than without AI assistance (Task 2), regardless of whether the model was certain or not. Also the acceptance rate of the AI's suggestions is lower.}
    \label{fig:uncert_uncert}
\end{figure*}
\FloatBarrier

\subsection{Potential for Blind Trust with Explanations}

Our study also identified a potential concern regarding the use of explanations, particularly in relation to fostering blind trust. Figures \ref{fig:acc_uncert} and \ref{fig:uncert_uncert} use color coding to represent user agreement with the model's suggestion (low agreement in blue, high agreement in red). These figures reveal a notable presence of "hot spots" where user agreement is high (green-yellow-orange-red) despite high model uncertainty. Table \ref{tab:user_tab} also shows that, on average, users' responses matched more often the model's prediction in Task 2 than in the other tasks. On the one hand, following the AI's recommendation lead to higher user accuracy scores, as discussed above. On the other hand, when zooming in only on the misclassified samples (accompanied by a matching wrong explanation in the four XAI groups), we report the following agreement rates: Control - 47.29\%, Control-Confidence - 52.31\%, CoProNN - 47.67\%, TCAV - 45.25\%, GradSim - 62.95\%, RepPoint - 69.1\%. This suggests that users may, in some cases, predominantly in the two example-based XAI groups, exhibit a tendency to blindly trust the model's suggestions, even when presented with explanations for demonstrably incorrect predictions.

\begin{table}
    \centering
    \caption{Aggregated user uncertainty scores and user-AI agreement rates for control tasks (1 and 3) and Task 2. The users' uncertainty was computed as described at the end of Section \ref{sec:exp_design}.}
    \begin{tabular}{c | c | c }
        \toprule
         & \textit{User Uncertainty} & \textit{User-AI Agreement} \\
         \midrule
        \textit{Tasks 1+3 (no AI help)} & 0.0359 & 0.5732 \\
        \midrule
        \textit{Task 2 (with AI assistance)} & 0.0151 & 0.7187 \\
        \bottomrule
    \end{tabular}
    \label{tab:user_tab}
\end{table}

\section{Conclusion}\label{sec:conc}

In this work, we proposed a novel HIL experiment design that allows to analyze the didactic effect of explanations, as well as correlations between the users' and the model's uncertainty. Apart from these new considerations, more traditional investigative points such as human-machine performance as a team or blind trust were also taken into account. We examined human-AI collaboration with explanations in image classification; nonetheless, our framework can be readily applied to any other machine learning task. 

We found that explanations considerably improved user performance during collaboration, especially when the AI was certain of its prediction. User uncertainty also decreased with explanations. However, our study also identified certain limitations. Explanations did not show notable benefits for long-term knowledge transfer and the specific explanation format had little to no impact on user accuracy. In line with previous work, our results also show that explanations tend to bias users responses to replicate the AI predictions, even when they are wrong; this finding highlights the need for well calibrated trust relationships in human-AI interactions in order to counteract blind trust in AI systems. We hope that the experimental paradigms quantifying the trust relationship in XAI developed in this study contribute to a better understanding of the trust relationship. 

Overall, our findings support the potential of human-AI collaboration with explanations to enhance performance and trust. Nevertheless, further research is needed to optimize the design of explanations for knowledge transfer and mitigate blind trust. Future work should explore how XAI can empower users not just to collaborate effectively, but also develop their own problem-solving skills. By striking a balance between trust and critical thinking, we believe that explanations can pave the path for a future of successful human-machine collaboration.


%
%
%
\bibliographystyle{splncs04}
\bibliography{egbib}
%

\end{document}